\documentclass[preprint,aps]{revtex4-1}

\usepackage{graphicx}

\begin{document}

\title{Electronic Structure Examination on the Topological Properties of CaMnSb$_2$ by Angle-Resolved Photoemission Spectroscopy}

\author{Hongtao Rong$^{1,2,\sharp}$, Liqin Zhou$^{1,2,\sharp}$, Junbao He$^{1,2,6,\sharp}$, Chunyao Song$^{1,2,\sharp}$, Jianwei Huang$^{1,2}$, Cheng Hu$^{1,2}$, Yu Xu$^{1,2}$, Yongqing Cai$^{1,2}$, Hao Chen$^{1,2}$, Cong Li$^{1,2}$, Qingyan Wang$^{1,2}$, Lin Zhao$^{1,2}$, Zhihai Zhu $^{1,2}$,  Guodong Liu$^{1,2,4}$, Zuyan Xu$^{3}$, Genfu Chen$^{1,2}$, Hongming Weng$^{1,2}$ and X. J. Zhou$^{1,2,4,5,*}$}

\affiliation{
\\$^{1}$National Lab for Superconductivity, Beijing National Laboratory for Condensed Matter Physics, Institute of Physics, Chinese Academy of Sciences, Beijing 100190, China
\\$^{2}$University of Chinese Academy of Sciences, Beijing 100049, China
\\$^{3}$Technical Institute of Physics and Chemistry, Chinese Academy of Sciences, Beijing 100190, China
\\$^{4}$Songshan Lake Materials Laboratory, Dongguan 523808, China
\\$^{5}$Beijing Academy of Quantum Information Sciences, Beijing 100193, China
\\$^{6}$Henan International Joint Laboratory of MXene Materials Microstructure, College of Physics and Electronic Engineering, Nanyang Normal University, Nanyang 473061, China
\\$^{*}$Corresponding author: XJZhou@iphy.ac.cn.
}

\date{\today}

\begin{abstract}
We have carried out detailed high resolution ARPES measurements and band structure calculations to study the electronic structure of CaMnSb$_2$. The observed Fermi surface mainly consists of one hole pocket around $\Gamma$ point and one tiny hole pocket at Y point. Strong spectral weight accumulation along the $\Gamma$-X direction is observed on the hole-like Fermi surface around $\Gamma$ point, suggesting strong anisotropy of the density of states along the Fermi surface. The tiny hole pocket at Y point originates from an anisotropic Dirac-like band with the crossing point of the linear bands lying $\sim$10 meV above the Fermi level. These observations are in a good agreement with the band structure calculations. In addition, we observe additional features along the $\Gamma$-Y line that cannot be accounted for by the band structure calculations. Our results provide important information in understanding and exploration of novel properties in CaMnSb$_2$ and related materials.
\end{abstract}


\maketitle

Topological materials have attracted much attention due to their novel quantum phenomena and physical properties as well as potential applications\cite{MZHasan,XLQi,ABansil,CFangTNL,BHYan,NPArmitage,YTokura}. Various types of topological materials have been discovered, including Quantum spin Hall states\cite{CLKaneQSHE,CLKaneZ2,BABernevig,MKonig}, three-dimensional topological insulators\cite{LFu,JEMoore,RRoy,DHsieh,YXia,YLChenBi2Te3}, Dirac semimetals\cite{PAMDirac,ZJWang,ZKLiuNa3Bi,AJLiang,ZKLiuCd3As2,MNeupaneCd3As2,HMYi}, Weyl semimetals\cite{XWan,SMHuang,HMWeng,SYXu,LXYang,BQLv}, and topological node-line semimetals\cite{AABurkov,LMSchoop,MNeupaneZSS,CFangTNL,YWangLST}. The Dirac materials present peculiar behaviours due to the presence of Dirac fermions. In Dirac materials, the low energy excitations are of relativistic nature and these quasiparticles obey Dirac equations, instead of the conventional Schr\"{o}dinger equation\cite{NPArmitage}. These Dirac fermions may give rise to exotic transport properties\cite{TOWehling,HDSong,HCWang}, such as anomalous quantum Hall effect\cite{KSNovoselov,YZhangQHE}, chiral anomaly\cite{SAParameswaran,JXiong}, ultrahigh mobility, giant magnetoresistance\cite{ZJWangC3A2,TLiangC3A2,JYYuanCd3As2,YFZhao} and Aharonov-Bohm oscillations\cite{LXWang}.

The AMnPn$_2$ (A = Ca, Sr, Ba, Eu or Yb; Pn= Sb or Bi) compounds have become a fertile playground for exploration of topological materials\cite{JParkSMB,JKWangSMB,YJJoSMB,YFeng,
JYLSMS,HMasudaEMB,AMZhangCSMB,JYLiuYMBNC,YShiomiEMB,SBorisenkoYMB,RYangCNMB,JBYu,JYLiuBMS}. The AMnPn$_2$ materials feature a common layered structure with an alternative stacking of A-Pn-A layers and MnPn$_4$ layers, as exemplified by CaMnSb$_2$ in Fig. 1a. In the A-Pn-A layer, the basic building block is the Pn sheet (Pn = Sb or Bi) in a regular or distorted square lattice, which can form highly anisotropic Dirac cones\cite{JParkSMB,GLeeAMB,YFeng} or Weyl cones\cite{SBorisenkoYMB}. In the MnPn$_4$ layers that are expected to be less conducting, Mn atoms are located at the center of edge-sharing tetrahedrons and the magnetic moments of the Mn atoms are ordered antiferromagnetically below the N\'{e}el temperature near room temperature. Dirac fermions in AMnPn$_2$ have been found to interplay with magnetism, leading to novel exotic properties, as has been demonstrated in Sr$_{1-y}$Mn$_{1-z}$Sb$_{2}$ (y, z $<$0.1)\cite{JYLSMS}, EuMnBi$_2$\cite{HMasudaEMB} and YbMnBi$_2$\cite{SBorisenkoYMB}. Compared with AMnSb$_2$, AMnBi$_2$ have stronger spin orbit coupling (SOC) due to heavier Bi atoms that opens a larger gap at Dirac nodes, leading to massive Dirac fermions\cite{JParkSMB,YFeng,GLeeAMB}. Replacement of Bi in AMnPn$_2$ materials by lighter elements such as Sb is expected to reduce the gap to achieve massless Dirac fermions.

CaMnSb$_2$ crystalizes in the Pnma space group\cite{JBHeCMS}. Compared with the crystal structure of AMnBi$_2$, instead of forming Bi square net plane, the Sb atoms in the Ca-Sb-Ca layer form zig-zag chains(Fig. 1b). The magnetotransport measurements suggest that CaMnSb$_2$ is a quasi-two-dimensional system that hosts nearly massless Dirac fermions\cite{JBHeCMS}. However, the infrared spectroscopic measurements indicate that CaMnSb$_2$ is a topologically trivial insulator\cite{ZYQiuCMS}. Angle-resolved photoemission spectroscopy (ARPES) is a powerful tool which can directly reveal the topological properties of CaMnSb$_{2}$, but so far no ARPES measurements have been reported for CaMnSb$_{2}$. In this paper, we have carried out detailed high resolution ARPES measurements and band structure calculations to study the electronic structure of CaMnSb$_2$. The observed Fermi surface mainly consists of one hole pocket around $\Gamma$ point and one tiny hole pocket at Y point. The tiny hole pocket at Y point originates from an anisotropic Dirac-like band that may play an important role in dictating the physical properties of CaMnSb$_2$.

High-quality single crystals of CaMnSb$_{2}$ were grown by flux method\cite{JBHeCMS}. ARPES measurements were performed at our lab-ARPES system equipped with the 6.994 eV vacuum-ultra-violet (VUV) laser, 21.218 eV Helium discharge lamp and Scienta DA30L electron energy analyzer\cite{GDLiu,XJZhou}. The energy resolutions for VUV laser and Helium discharge lamp ARPES measurements were set at $\sim$1 meV and 20 meV, respectively. The angular resolution was $\sim0.3^{\circ}$. The Fermi level is referenced by measuring on the Fermi edge of a clean polycrystalline gold that is electrically connected to the sample. All the samples were cleaved {\it in situ} and measured in ultrahigh vacuum with a base pressure better than $5\times10^{-11}$ mbar.

The electronic structure calculation of CaMnSb$_{2}$ based on the density functional theory (DFT) were performed by using the Vienna ab initio simulation package (VASP)\cite{GKresse}. The generalized gradient approximation (GGA) in the Perdew-Burke-Ernzerhof (PBE) type was selected to describe the exchange-correlation function\cite{JPPerdew}, and the modified Becke-Johnson (mBJ) potential was used to achieve accurate band gap calculations\cite{ADBecke,FTran}. The cutoff energy was set to 500 eV and the Brillouin zone (BZ) integration was sampled by $3 \times 11 \times 11$ k mesh. Spin-orbital coupling (SOC) was taken into account. The tight-binding model of CaMnSb$_{2}$ was constructed by the Wannier90 with Sb 5$p$ orbitals and Mn 3$d$ orbitals, which are based on the maximally-localized Wannier functions (MLWF)\cite{AAMostofi}. The bulk Fermi surfaces and projected surface states of CaMnSb$_{2}$ were calculated by the WannierTools package\cite{QSWu}.

Figure 1 shows the crystal structure and the calculated band structures of CaMnSb$_{2}$. Fig. 1a shows the crystal structure of CaMnSb$_{2}$ which consists of an alternate stacking of MnSb$_{4}$ layers and Ca-Sb-Ca layers along the {\it a} axis. In the Ca-Sb-Ca layer, the central Sb sheet is sandwiched in between two Ca layers; the Sb atoms in the Sb sheet form zig-zag chains with the chain direction along the {\it b} axis (Fig. 1b). It has a space group of Pnma and an orthorhombic crystal structure with the lattice constants of a=22.09 {\AA}, b=4.32 {\AA} and c=4.35 {\AA}. The corresponding Brillouin zone is shown in Fig. 1c. Fig. 1d shows the overall calculated band structure of CaMnSb$_{2}$ without considering SOC. The low energy band structure near $\Gamma$ point mainly consists of two bands. The band structure near Y point shows a Dirac cone-like structure. However, even without considering SOC, a small gap already opens at the Dirac point and the position of the Dirac point deviates slightly from Y point along the Y-M direction. When SOC is taken into account, it has little effect on the bands near $\Gamma$ point but dramatically changes the band structure near Y point. The gap at the Dirac point is greatly enhanced up to $\sim$180 meV. In the mean time, the band top of the lower branch shifts to Y point along the Y-M direction. In order to get accurate band gap, we carried out band structure calculations of CaMnSb$_{2}$ by using the modified Becke-Johnson (mBJ) potential\cite{ADBecke,FTran}, as shown in Fig. 1g and 1h for $k_{z}=0$ and $k_{z}=\pi/a$, respectively. The inclusion of mBJ potential alters the relative energy position of the bands near $\Gamma$ and Y points. The initial bands at high binding energy are pushed further to even higher binding energy. It is interesting to note that the two main bands at $\Gamma$ and Y points for $k_{z}=0$ (Fig. 1g) become nearly degenerate when $k_{z}$ moves to $\pi/a$ (Fig. 1h). We also analysed the orbital contributions to the low energy bands, as shown in Fig. 1f. The bands near the Fermi level originate mainly from the Sb 5{\it p} orbitals. Moreover, the bands near $\Gamma$ point mainly come from the Sb 5{\it p$_{x}$} orbitals while the bands near Y point along Y-$\Gamma$ and Y-M directions are mainly from Sb 5{\it p$_{z}$} and 5{\it p$_{y}$} orbitals, respectively.

Figure 2 shows CaMnSb$_{2}$ Fermi surface and constant energy contours at different binding energies measured by using 21.218 eV photon energy. In order to obtain complete electronic structures, we carried out ARPES measurements under two distinct polarization geometries. When the electric vector of the light is along the $\Gamma$-Y direction (Fig. 2a), the observed Fermi surface consists mainly of a circular sheet around $\Gamma$ and a strong spot at Y (leftmost panel in Fig. 2a); no feature at X is observed. With increasing binding energy, the central circular sheet increases in area, indicating the corresponding Fermi surface is hole-like. The area of the hole-like pocket is estimated to be $\sim$0.1 {\AA}$^{-2}$. In the meantime, the strong spot at Y also increases in its area up to a binding energy of $\sim$200 meV; a triangle-like feature develops at higher binding energy and it connects Y point and the central circular sheet (rightmost panel in Fig. 2a). When the electric vector of the light is switched to be along the $\Gamma$-X direction (Fig. 2b), a similar hole-like Fermi surface is observed around $\Gamma$. However, in this case, the strong spot appears at X but is absent at Y.

The calculated Fermi surface and constant energy contours of CaMnSb$_{2}$ (Fig. 2c) show a good agreement with the measured results in Fig. 2a. The zig-zag Sb chains in CaMnSb$_{2}$ give rise to the highly anisotropic electronic structure that leads to the observation of tiny pocket at Y but its absence at X. Since the bands near Y point are mainly composed of Sb 5{\it p$_{z}$} and 5{\it p$_{y}$} orbitals (Fig. 1f), by considering the matrix element effects in photoemission process, the states are allowed under the used polarization geometry (Fig. 2a). The observation of the tiny pocket around X point in Fig. 2b can be explained by the presence of two orthogonal domains in the orthorhombic {\it bc} plane of CaMnSb$_{2}$. The twin structure has been revealed previously in the related sister compounds SrMnSb$_{2}$\cite{SVRamankuttySMS} and BaMnSb$_{2}$\cite{JYLiuBMS}. Our measured Fermi surface of CaMnSb$_{2}$ is also consistent with that of single-domain SrMnSb$_{2}$\cite{SVRamankuttySMS}.

Now we zoom in onto the detailed electronic structure around Y point in CaMnSb$_{2}$. Fig. 3a shows the constant energy contours around Y point at different binding energies. With increasing binding energy, the tiny hole pocket around Y point gradually goes from ellipse-like to rectangle-like shape. Fig. 3b shows the band evolution along the different momentum cuts around Y. Dirac cone-like bands centered around Y (cut 5) are observed from cut 3 to cut 7 which contribute to the formation of the tiny hole pocket around Y point. We also observe additional bands around k$_{x}$= $\pm$0.2 ($\pi/b$) at the binding energy higher than 250 meV (cuts 1, 2, 8 and 9 in Fig. 3b). These bands contribute to the triangle-like feature that connects Y point and the central circular sheet in the constant energy contour (rightmost panel in Fig. 2a). The Dirac-like band along cut 5 (Y-M direction crossing Y point) exhibits steep linear dispersion, as seen in Fig. 3c. From the corresponding momentum distribution curves (MDCs) at different binding energies (Fig. 3d), we find that the two peaks in the MDCs at high binding energies gradually evolve into a single peak at $\sim$10 meV above the Fermi level. This indicates the crossing point of the two branches of the Dirac-like band lies around 10 meV above the Fermi level. The linear fitting of the MDC dispersions in Fig. 3d yields a Fermi velocity of $\sim$4.3 eV$\cdot${\AA}. Fig. 3e shows the corresponding photoemission spectra (energy distribution curves, EDCs) of Fig. 3c. A peak can be observed for the EDC at Y point ($k_{T}$ as marked in Fig. 3c). The band structure along $\Gamma$-Y direction crossing Y point (cut 10 in Fig. 3a) is shown in Fig. 3f. It also exhibits a Dirac-like band structure with a less steep dispersion than the one in Fig. 3c, giving a Fermi velocity of $\sim$2.2 eV$\cdot${\AA}.  Combining the results in Fig. 3a, Fig. 3c and Fig. 3f, we arrive at a three-dimensional pyramid structure that represents an anisotropic Dirac cone in momentum space near Y point.

Next, we investigate the detailed electronic structure of CaMnSb$_{2}$ around $\Gamma$ point. In addition to the measurements using the Helium-discharge lamp (h$\nu$=21.218 eV) (Fig. 2a), we also carried out high resolution ARPES measurements using a laser (h$\nu$=6.994 eV). Fig. 4a and 4b show Fermi surface mappings measured under two different photon polarizations using the 6.994 eV laser. The hole-like Fermi surface around $\Gamma$, seen in Fig. 2a and 2b with 21.218 eV Helium lamp, is more clearly observed by using the 6.994 eV laser. The area of the hole-like Fermi pocket around $\Gamma$ point in Fig. 4a and 4b is estimated to be about 0.17 {\AA}$^{-2}$ which is slightly different from that measured by Helium lamp ($\sim$0.1 {\AA}$^{-2}$, Fig. 2a and 2b), reflecting the k$_{z}$ dependence of the electronic structure. It is interesting to note that the spectral weight along the hole-like Fermi surface in Fig. 4a and 4b exhibits a strong variation. Under the horizontal light polarization in Fig. 4a, two strong spots are observed on the Fermi surface along the horizontal $\Gamma$-X direction while the spectral weight along the vertical $\Gamma$-Y direction is apparently suppressed. When the light polarization is switched to the vertical direction in Fig. 4b, the two strong spots on the Fermi surface also change to the vertical $\Gamma$-Y direction while the spectral weight along the horizontal $\Gamma$-X direction is suppressed.

In order to understand the origin of the strong intensity variation along the Fermi surface(Fig. 4a and 4b), we present in Fig. 4c the band structures along different momentum cuts crossing the $\Gamma$ point from the data in Fig. 4a. Fig. 4d shows the MDCs at the Fermi level from the bands in Fig. 4c. Indeed, the intensity of the hole-like bands drops dramatically when the momentum cut changes from the horizontal $\Gamma$-X direction to the vertical $\Gamma$-Y direction, as seen directly from the band intensity in Fig. 4c and the MDC peak intensity in Fig. 4d. Fig. 4e shows quantitatively the spectral weight variation along the Fermi surface by plotting the MDC peak intensity as a function of the Fermi surface angle $\theta$. When the light polarization E is horizontal (Fig. 4a), the spectral weight along the hole-like Fermi surface peaks sharply at $\theta$=0 and 180 degrees (blue line in Fig. 4e). When the light polarization E is vertical (Fig. 4b), the sharp spectral weight peak along the hole-like Fermi surface changes into $\theta$=90 and 270 degrees (red line in Fig. 4e). Fig. 4f shows EDCs along the $\Gamma$-X momentum cut. The EDCs on the hole-like Fermi surface near the strong spot region show broad peaks rather than sharp quasiparticle peaks, which indicates that the electronic states along the hole-like Fermi surface experience strong scattering.

The first issue to address is whether the strong spot feature is a part of the hole-like Fermi surface or it is a new feature in addition to the hole-like Fermi surface. Careful examination of the band structures near the strong spot region indicates that there is only a single hole-like band; no indication of two coexisting features is observed (Fig. 4c). This indicates that the strong spot feature is a part of the hole-like Fermi surface.  The second issue concerns whether the strong spots originate from the photoemission matrix element effect. The band structure calculations have shown that the entire hole-like Fermi surface is composed of Sb {\it p$_{x}$} orbitals, as schematically shown in the inset of Fig. 4e. When the light polarization is horizontal, as in the case of Fig. 4a, all the states along the hole-like Fermi surface are allowed according to the matrix element effect analysis. If the density of states along the Fermi surface is uniform, one would expect to observe uniform spectral weight distribution along the hole-like Fermi surface that is not consistent with the observation of two strong spots on the Fermi surface. These results suggest that the density of states along the hole-like Fermi surface is strongly anisotropic; there is a strong accumulation of the density of states peaked along the $\Gamma$-X direction. Such a confinement of strong density of states at two particular points along the Fermi surface is unusual and its effect on the physical properties is interesting to investigate. The polarization dependence of the measured Fermi surface mappings in Fig. 4a and 4b can be understood when the coexisting two kinds of domains (represented by domains A and B) are considered. For a given polarization in Fig. 4a, from the photoemission matrix element effect analysis the electronic states along the entire hole-like Fermi surface are allowed for the domain A but are fully suppressed for the domain B that is 90 degree rotated with respect to the domain A; the measured Fermi surface mainly comes from one domain A. By the similar analysis, the rotated pattern in Fig. 4b can be understood when the light polarization is 90 degree rotated because the measured Fermi surface in this case comes mainly from another domain B.

In addition to the hole-like Fermi pocket, we also observed another band around $\Gamma$ point, as can be seen in Fig. 4c. In Fig. 5, we focus on this feature by analysing the data taken with both the 6.994 eV laser and the 21.218 eV Helium lamp. Fig. 5a-5c shows Fermi surface and constant energy contours at different binding energies measured by using 6.994 eV photon energy. Three typical bands measured along the momentum cuts parallel to $\Gamma$-X direction are shown in Fig. 5d. The central band is observed for all three momentum cuts at different k$_{y}$s. The MDC analyses in Fig. 5e indicate that this central band is weak at the Fermi level and low binding energy. It gets stronger with increasing binding energy and becomes dominant at high binding energy. The central band can be more clearly visualized in the second derivative images in Fig. 5f; it appears to be composed of two nearly vertical bands on the two sides of the momenta along the $\Gamma$-Y line. The persistence of this band for the momentum cuts at different k$_{y}$s results in a strong intensity patch around the $\Gamma$-Y line at a high binding energy as shown in Fig. 5c. This band is also observed in the measurements using 21.218 eV Helium lamp (Fig. 5g, 5h and 5i). In this case, we find that the signal of the hole-like band sits on top of a strong background formed by the electronic states around $\Gamma$ point (Fig. 5h). The existence of the band at $\Gamma$ and along $\Gamma$-Y line is clear. But the band structure calculations (Fig. 1g) do not reproduce the corresponding band at $\Gamma$ point within 0.5 eV energy range below the Fermi level. In order to find out whether this band stems from surface states, we also calculated the surface states in CaMnSb$_{2}$ as shown in Fig. 5j. We find that the new band observed at $\Gamma$ point cannot be attributed to the surface states either. The origin of this additional band needs further investigations.

Our present ARPES measurements on CaMnSb$_{2}$ make it possible to compare with SrMnSb$_{2}$ and BaMnSb$_{2}$ to investigate the effect of replacing the alkali earth metal elements in AMnSb$_{2}$ (A=Ca, Sr, Ba) on the electronic structures. BaMnSb$_{2}$ crystalizes in the I2mm space group and from the ARPES measurement, two small hole pockets near the Y point are observed\cite{JYLiuBMS,HSakaiBMS}. On the other hand, CaMnSb$_{2}$ crystalizes in the Pnma space group and from our ARPES measurements, there is only one hole pocket at the Y point. Therefore, there is a significant difference between CaMnSb$_{2}$ and BaMnSb$_{2}$ in both crystal structure and electronic structure. Although the band structure of CaMnSb$_{2}$ is similar to SrMnSb$_{2}$ \cite{SVRamankuttySMS} in some aspects, from our high resolution ARPES measurements, we have uncovered new observations in CaMnSb$_{2}$ that are distinct from SrMnSb$_{2}$. First, the strong spectral weight variation along the hole-like Fermi surface around $\Gamma$ in CaMnSb$_{2}$ that is not observed in SrMnSb$_{2}$\cite{SVRamankuttySMS}. Second, we revealed an additional band around $\Gamma$ in CaMnSb$_{2}$ that is not observed in SrMnSb$_{2}$\cite{SVRamankuttySMS}. Third, we find that the tiny hole pocket at the Y point originates from an anisotropic Dirac-like band with the crossing point of the linear bands lying $\sim$10 meV above the Fermi level in CaMnSb$_{2}$ while it is $\sim$200 meV above the Fermi level in SrMnSb$_{2}$\cite{SVRamankuttySMS}. This indicates that the Dirac-like band in CaMnSb$_{2}$ is unique in generating exotic physical properties.

It has been found that the electronic structures and topological properties of AMnPn$_2$ are intimately connected with their crystal structures. So far four distinct crystal structures have been identified in AMnPn$_{2}$. In Fig. 6, we summarize crystal structures and electronic structures of AMnPn$_2$ (A = Ca, Sr, Ba, Eu or Yb; Pn= Sb or Bi). For CaMnBi$_2$\cite{YFeng}, YbMnBi$_2$\cite{SBorisenkoYMB} and YbMnSb$_2$\cite{RKealhoferYMS} in P4/nmm space group, the Fermi surface mainly consists of a large diamond-like Fermi surface connecting four equivalent X points in the first Brillouin zone. Dirac cones with a small gap opening at the Dirac point are formed along the diamond-like Fermi surface, with the energy position of the Dirac points varying slightly in the momentum space. For SrMnBi$_2$\cite{JParkSMB,YFeng,LLJiaSMB}, BaMnBi$_2$\cite{HRyuBMB} and EuMnBi$_2$\cite{SBorisenkoYMB} in I4/mmm space group, the Fermi surface shows four separated crescent-moon-like segments along $\Gamma$-M direction. In this case, Dirac cones are also formed along a line, but the gap opening at the Dirac point is small along the $\Gamma$-M and gets much large along the $\Gamma$-X direction. For BaMnSb$_2$\cite{JYLiuBMS,HSakaiBMS} in I2mm space group, its Fermi surface is composed mainly of a large hole-like pocket around $\Gamma$ point and two Dirac cone-like hole pockets around Y point. For SrMnSb$_2$\cite{SVRamankuttySMS}, EuMnSb$_{2}$\cite{JRSohEMS} and CaMnSb$_2$ in Pnma space group, the Fermi surface consists of a hole-like pocket around $\Gamma$ point, but only one Dirac cone-like hole pocket at Y point.

The Fermi surface of CaMnSb$_2$ mainly consists of one hole pocket around $\Gamma$ point and one tiny Dirac cone-like hole pocket at Y point. We have found that the electronic states along the hole-like Fermi surface experience strong scattering because the measured EDCs are broad (Fig. 4f). On the other hand, the electrons on the tiny hole pocket at Y point experience much less scattering because a coherence peak can be observed (Fig. 3e). Therefore, the tiny hole pocket at Y point plays more important role in dictating the physical properties of CaMnSb$_2$. Quantum oscillation measurement detected a Fermi pocket with an area of 0.0008{\AA}$^{-2}$\cite{JBHeCMS} which agrees well with the measured area of the tiny hole pocket at Y ($\sim$0.0007 {\AA}$^{-2}$). But another Fermi pocket with an area of 0.0015{\AA}$^{-2}$ in quantum oscillation measurement\cite{JBHeCMS} has no clear correspondence in our ARPES measurements. On the other hand, the hole-like Fermi pocket around $\Gamma$ in our ARPES measurements is not detected in the quantum oscillation measurement\cite{JBHeCMS}. This may be related to the strong electron scattering on this hole-like Fermi pocket. 

In summary, we have carried out detailed high resolution ARPES measurements and band structure calculations to study the electronic structure of CaMnSb$_2$. The observed Fermi surface mainly consists of one hole pocket around $\Gamma$ point and one tiny hole pocket at Y point. We find strong spectral weight accumulation along the $\Gamma$-X direction on the hole-like Fermi surface around $\Gamma$ point, suggesting strong anisotropy of the density of states along the Fermi surface. The tiny hole pocket at Y point originates from an anisotropic Dirac-like band with the crossing point of the linear bands lying $\sim$10 meV above the Fermi level. These observations are in a good agreement with the band structure calculations. In addition, we also observe additional features along the $\Gamma$-Y line that cannot be accounted for by the band structure calculations. Our results indicate that the Dirac-like structure at Y point may play an important role in dictating the physical properties of CaMnSb$_2$. They provide important information in understanding and exploration of novel properties in CaMnSb$_2$ and related materials.

\vspace{3mm}

\textbf{Acknowledgments}
This work is supported by the National Key Research and Development Program of China
(Nos. 2016YFA0300600, 2018YFA0305602, 2016YFA0300300 and 2017YFA0302900), the National Natural Science Foundation of China (Nos. 11974404, 11888101, 11922414 and 11404175), the Strategic Priority Research Program (B) of the Chinese Academy of Sciences (Nos. XDB33000000 and  XDB25000000), the Youth Innovation Promotion Association of CAS (No. 2017013)and the Natural Science Foundation of Henan Province (Nos. 182300410274 and 202300410296). The theoretical calculations is supported by the National Natural Science Foundation of China (Grant Nos. 11674369, 11865019 and 11925408), the Beijing Natural Science Foundation (Grant No. Z180008), Beijing Municipal Science and Technology Commission (Grant No. Z191100007219013), the National Key Research and Development Program of China (Grant Nos. 2016YFA0300600 and 2018YFA0305700), the K. C. Wong Education Foundation (Grant No. GJTD-2018-01) and the Strategic Priority Research Program of Chinese Academy of Sciences (Grant No. XDB33000000)

\vspace{3mm}

\textbf{Author contributions}
  H.T.R, L.Q.Z, J.B.H and C.Y.S contribute equally to this work. X.J.Z. and H.T.R. proposed and designed the research. H.T.R. and Y.X. carried out the ARPES experiments. L.Q.Z. and H.M.W. contributed in the band structure calculations. J.B.H. and G.F.C. prepared the samples; H.T.R., C.Y.S., J.W.H., C.H., Y.X., Y.Q.C., H.C., C.L., Q.Y.W., L.Z., Z.H.Z., G.D.L. Z.Y.X. and X.J.Z. contributed to the development and maintenance of Laser-ARPES systems, the data analysis and the related software development. X.J.Z. and H.T.R. wrote the paper. All authors participated in discussions and comments on the paper.

\vspace{3mm}
\textbf{Additional information}

\vspace{3mm}
\textbf{Competing financial interests}:
 The authors declare no competing financial interests.

\newpage

\begin{figure*}[t]
\begin{center}
\includegraphics[width=0.99\columnwidth,angle=0]{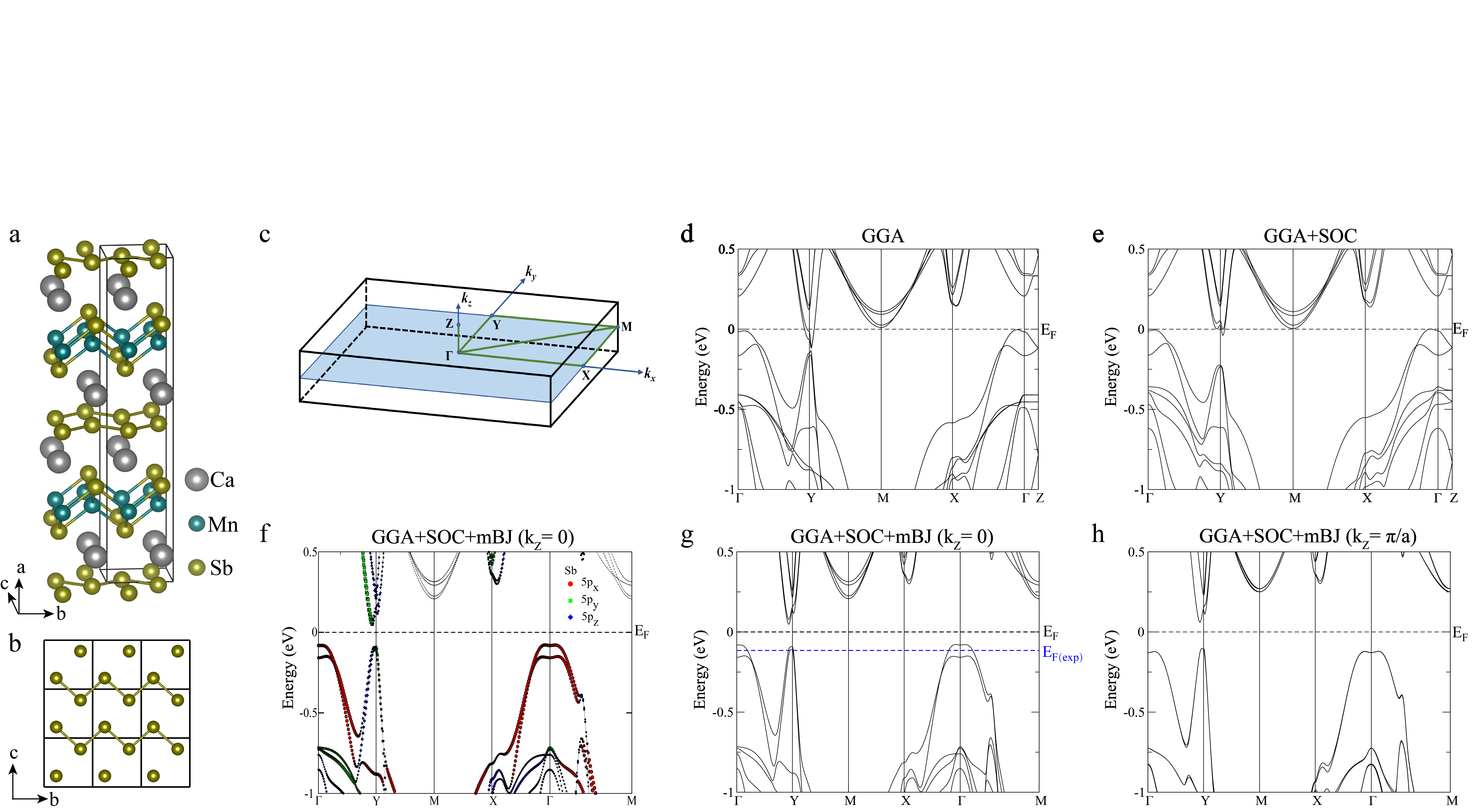}
\end{center}
\begin{center}
\caption{{\bf Crystal structure and calculated band structures of CaMnSb$_{2}$.} (a) Crystal structure of CaMnSb$_{2}$. The unit cell is indicated by black lines. It crystalizes in an orthorhombic structure (space group: Pnma) with lattice constants a=22.09 {\AA}, b=4.32 {\AA} and c=4.35 {\AA}. (b) Top view of the central Sb layer which consists of zig-zag chainlike structure along the b axis. (c) Three dimensional Brillouin zone of CaMnSb$_{2}$. Green lines indicate high symmetry directions. (d,e) Calculated band structure of CaMnSb$_{2}$ without (d) and with (e) spin-orbit coupling at $k_{z}=0$. (f) Calculated band structure of CaMnSb$_{2}$ by considering spin-orbit coupling and mBJ potential at $k_{z}=0$. The bands near the Fermi level originate mainly from Sb 5$p_{x}$ (red circles), 5$p_{y}$ (green squares) and 5$p_{z}$ (blue diamonds) orbitals. (g,h) Calculated band structure of CaMnSb$_{2}$ by considering spin-orbit coupling and mBJ potential at $k_{z}=0$ (g) and $k_{z}=\pi/a$ (h).
}
\end{center}
\end{figure*}

\begin{figure*}[t]
\begin{center}
\includegraphics[width=0.99\columnwidth,angle=0]{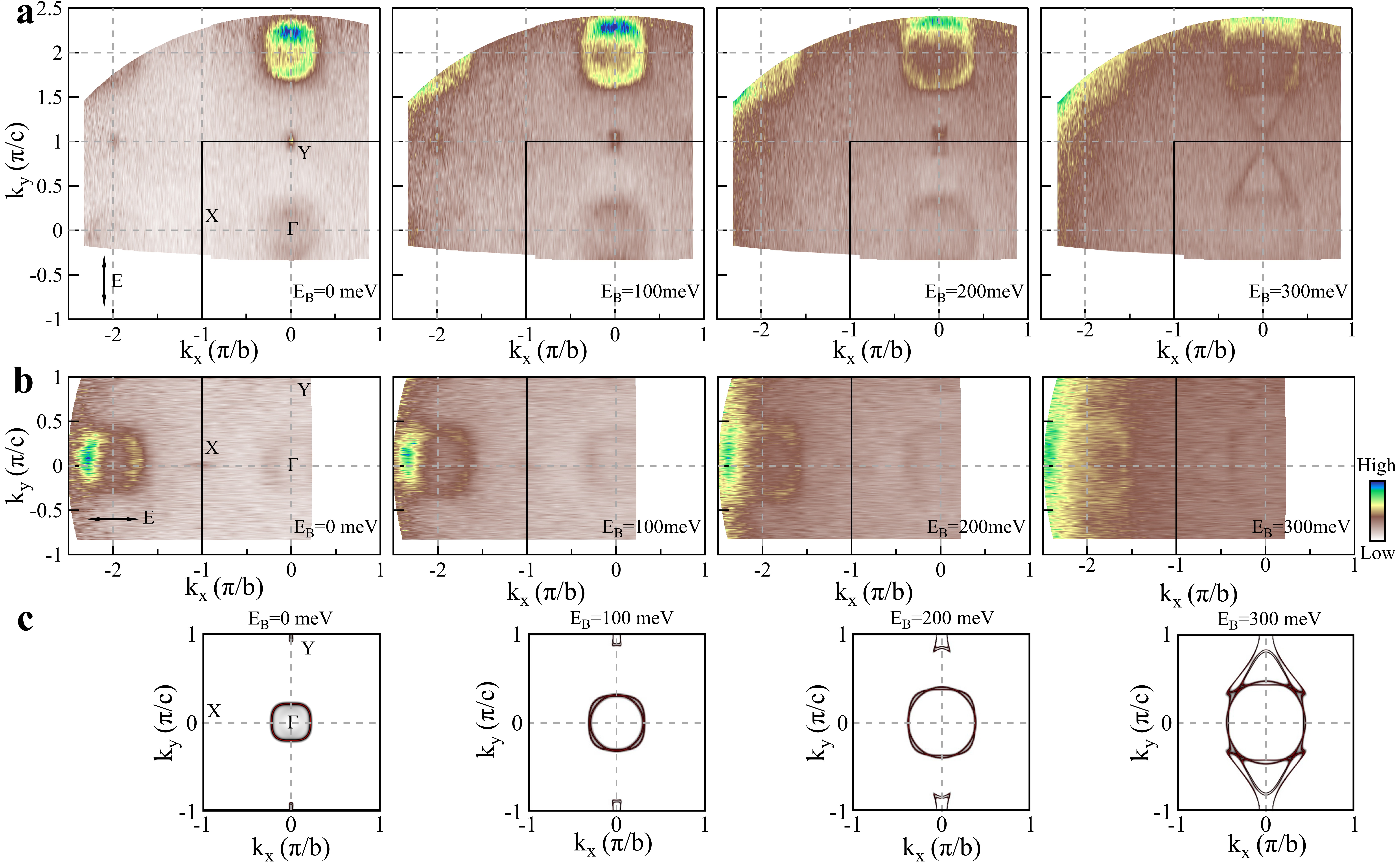}
\end{center}
\begin{center}
\caption{{\bf Measured and calculated Fermi surface mapping and constant energy contours of CaMnSb$_{2}$.} (a) Constant energy contours of CaMnSb$_{2}$ at binding energies of 0 meV, 100 meV, 200meV and 300 meV measured at 30 K with 21.218 eV Helium lamp. The major E vector is along $k_{y}$ direction, as marked by the arrow at the bottom-left inset of the left-most panel. (b) Same as (a). The major E vector is along $k_{x}$ direction, as marked by the arrow at the bottom-left inset of the left-most panel. All these constant energy contours in (a) and (b) are obtained by integrating the photoemission spectral weight over a [-20,20] meV energy window with respect to the corresponding binding energy. The first Brillouin zone is indicated by the black squares. (c) Calculated constant energy contours at binding energies of 0 meV, 100 meV, 200meV and 300 meV for $k_{z}=0$. An experimental Fermi level, E$_{F(exp)}$ is taken which is 0.12 eV below the Fermi level in the calculation, as shown in Fig. 1g.
}
\end{center}
\end{figure*}

\begin{figure*}[tbp]
\begin{center}
\includegraphics[width=0.99\columnwidth,angle=0]{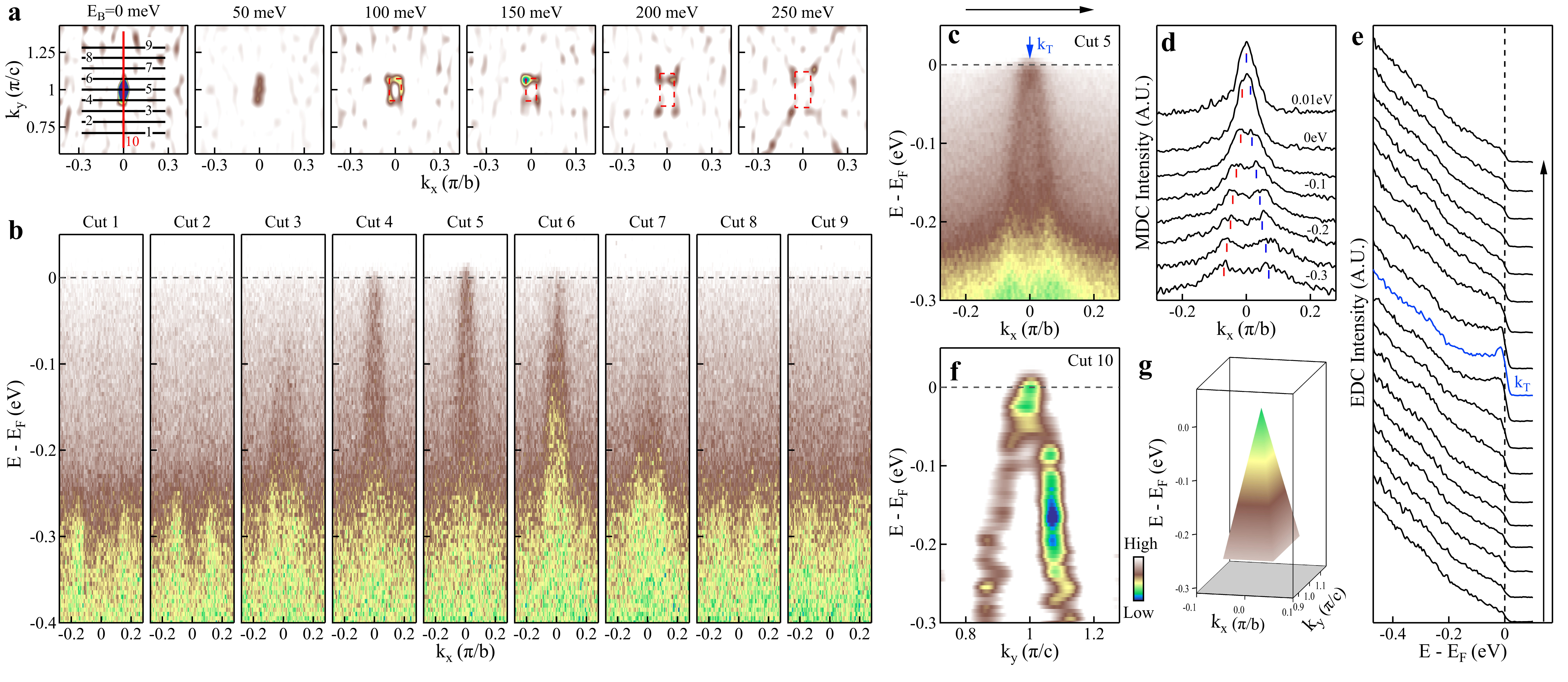}
\end{center}
\begin{center}
\caption{{\bf Dirac cone-like structure around Y point in CaMnSb$_2$ measured with 21.218 eV photon energy at 30 K.} (a) The second derivative constant energy contours around Y point obtained by integrating the photoemission spectral weight over a [-20,20] meV energy window with respect to the binding energy of 0 meV, 50meV, 100 meV, 150 meV, 200 meV and 250 meV. (b) Band structure measured along various momentum cuts. The location of the momentum cuts, cut 1 to cut 9, is marked by black lines in the left-most panel of (a). (c) Detailed band structure taken along the momentum cut 5 that is along YM direction crossing Y point. (d) Momentum distribution curves (MDCs) of the photoemission image in (c) at different binding energies. (e) Photoemission spectra (Energy dispersion curves, EDCs) at different momenta from photoemission image in (c). The blue line represents the EDC at Y point (marked by blue arrow and $k_{T}$ in (c)). (f) Band structure taken along the momentum cut 10. The location of cut 10 is marked by red line in the left-most panel of (a) that crosses Y point. The image is the second derivative of the original data with respect to the momentum. (g) Three-dimensional schematic of the Dirac cone-like structure at Y point.
}
\end{center}
\end{figure*}

\begin{figure*}[tbp]
\begin{center}
\includegraphics[width=0.99\columnwidth,angle=0]{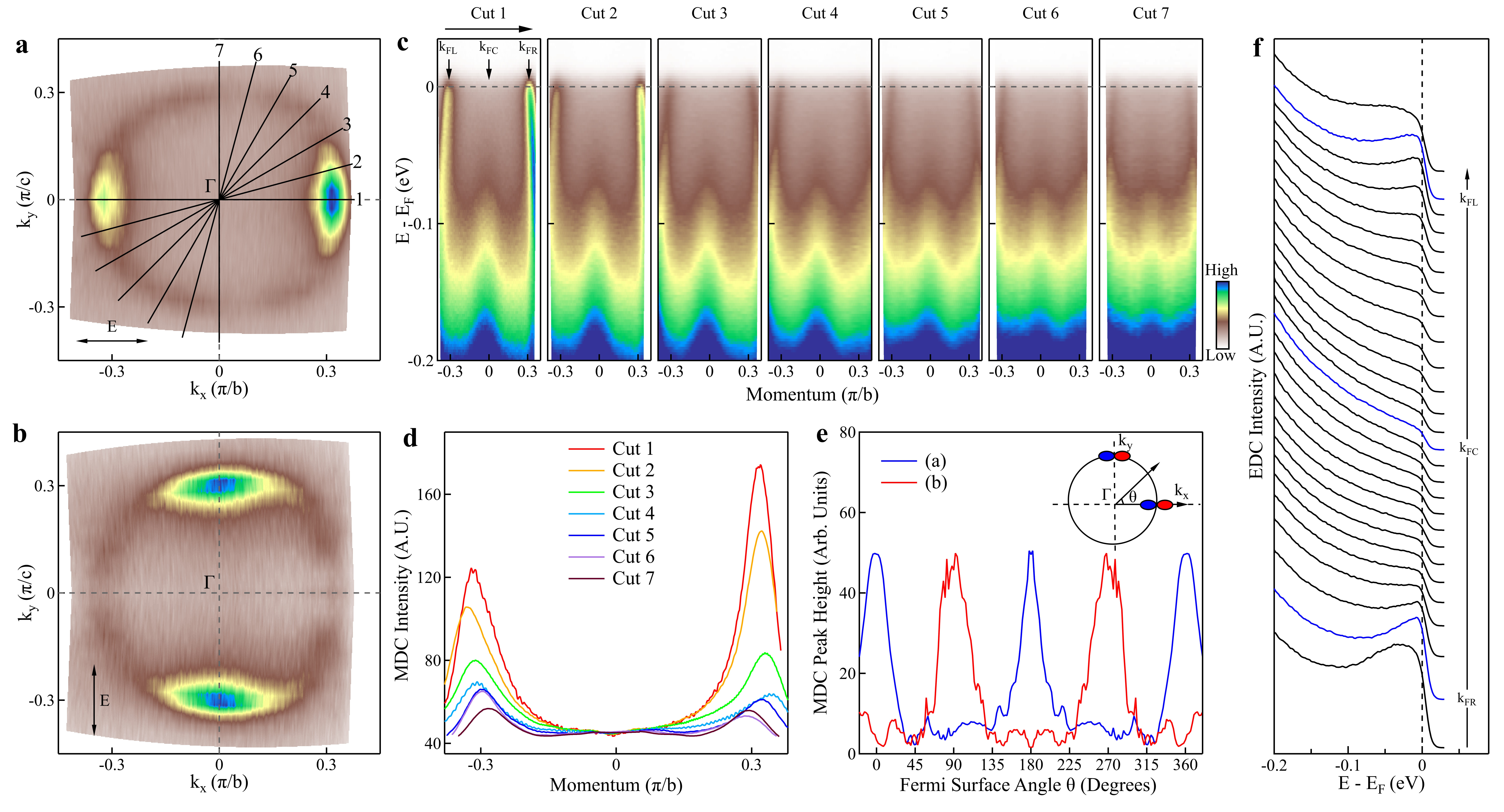}
\end{center}
\begin{center}
\caption{{\bf Detailed hole-like Fermi surface and band structures of CaMnSb$_2$ around $\Gamma$ point measured at 30 K.} (a) Fermi surface mapping measured with 6.994 eV laser by integrating the photoemission spectral weight over a [-10,10] meV energy window. The E vector is along $k_{x}$ direction, as marked by the arrow at the bottom-left corner. (b) Same as (a) but the E vector is along $k_{y}$ direction, as marked by the arrow at the bottom-left corner. (c) Band structure measured along various momentum cuts; the location of the momentum cuts is marked by black lines in (a). (d) MDCs at the Fermi level from the bands in (c) measured along various momentum cuts. (e) The angle-dependent MDC peak height along the Fermi surface in (a)(blue line) and (b)(red line) as a function of the Fermi surface angle. The Fermi surface angle $\theta$ is defined in the up-right inset. The entire hole-like Fermi pocket is composed of Sb {\it $p_{x}$} orbital. (f) EDCs at different momenta from the Cut 1 photoemission image (leftmost panel in (c)). The EDCs at the Fermi momenta on the left side ($k_{FL}$) and right side ($k_{FR}$) of the band, and at $\Gamma$ point($k_{FC}$) are marked by blue lines.
}
\end{center}
\end{figure*}

\begin{figure*}[tbp]
\begin{center}
\includegraphics[width=0.99\columnwidth,angle=0]{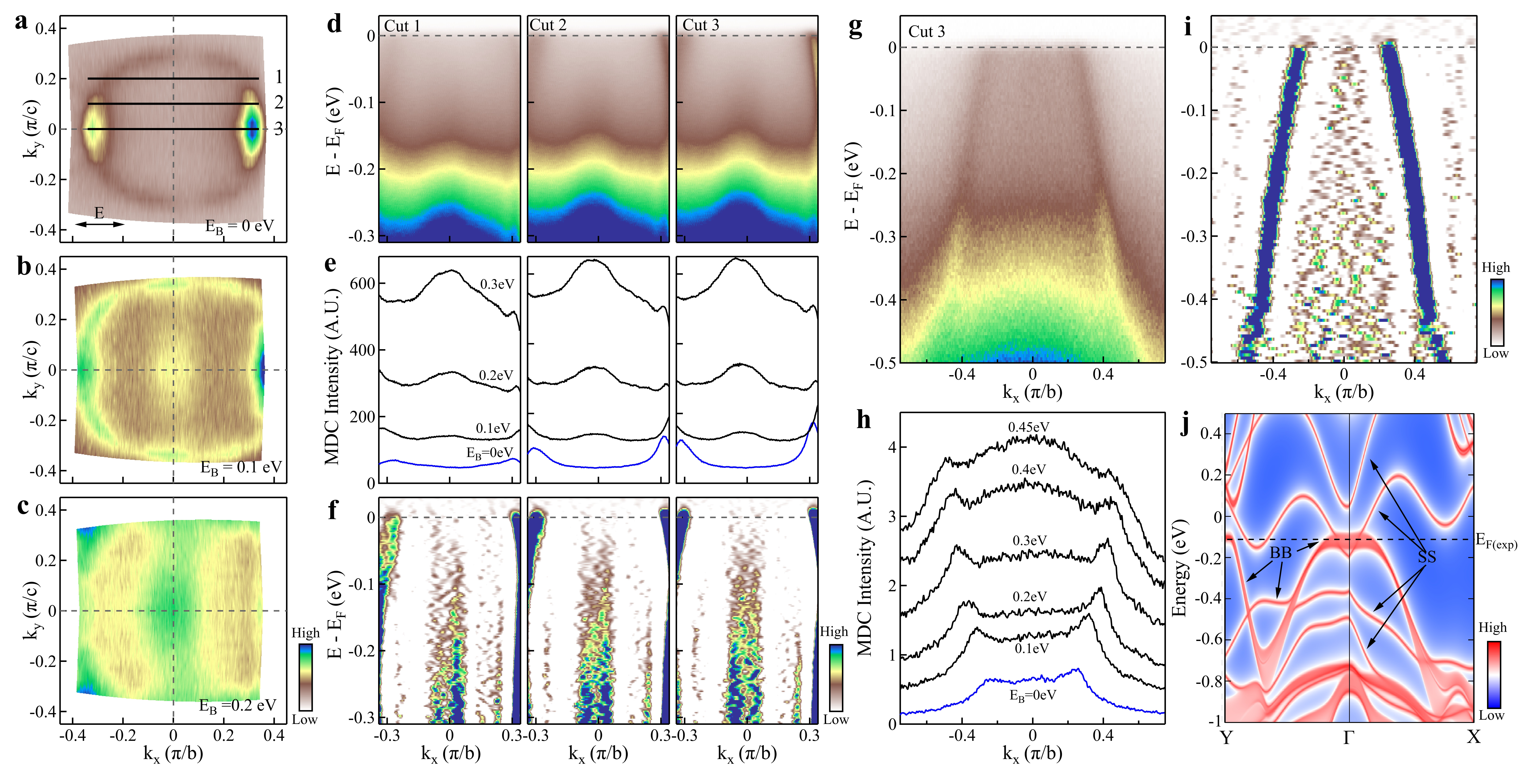}
\end{center}
\begin{center}
\caption{{\bf Detailed electronic structure of CaMnSb$_2$ around $\Gamma$ point measured at 30 K.} (a-c) Constant energy contours around $\Gamma$ obtained by integrating the photoemission spectral weight over a [-10,10] meV energy window with respect to 0 eV (a), 0.1 eV (b), 0.2 eV (c) binding energies. The data are the same as in Fig. 4a measured with 6.994 eV laser and the E vector is along $k_{x}$ direction, as marked by the arrow at the bottom-left inset of (a). (d) Band structure measured along various momentum cuts; the location of the momentum cuts is marked by black lines in (a). (e) MDCs at the different binding energies from the bands in (d) measured along various momentum cuts. (f) The corresponding second derivative images with respect to momentum of the band structures in (d). (g) Band structure measured with 21.218 eV Helium lamp along the horizontal $\Gamma$-X direction (same as cut 3 marked by the black line in (a)). (h) MDCs at the different binding energies from the band structure in (g). (i) The corresponding second derivative image with respect to momentum of the band structure in (g). (j) DFT-projected bulk bands and surface bands of the Sb-terminated (100) surface along the Y-$\Gamma$-X directions.
}
\end{center}
\end{figure*}

\begin{figure*}[tbp]
\begin{center}
\includegraphics[width=0.99\columnwidth,angle=0]{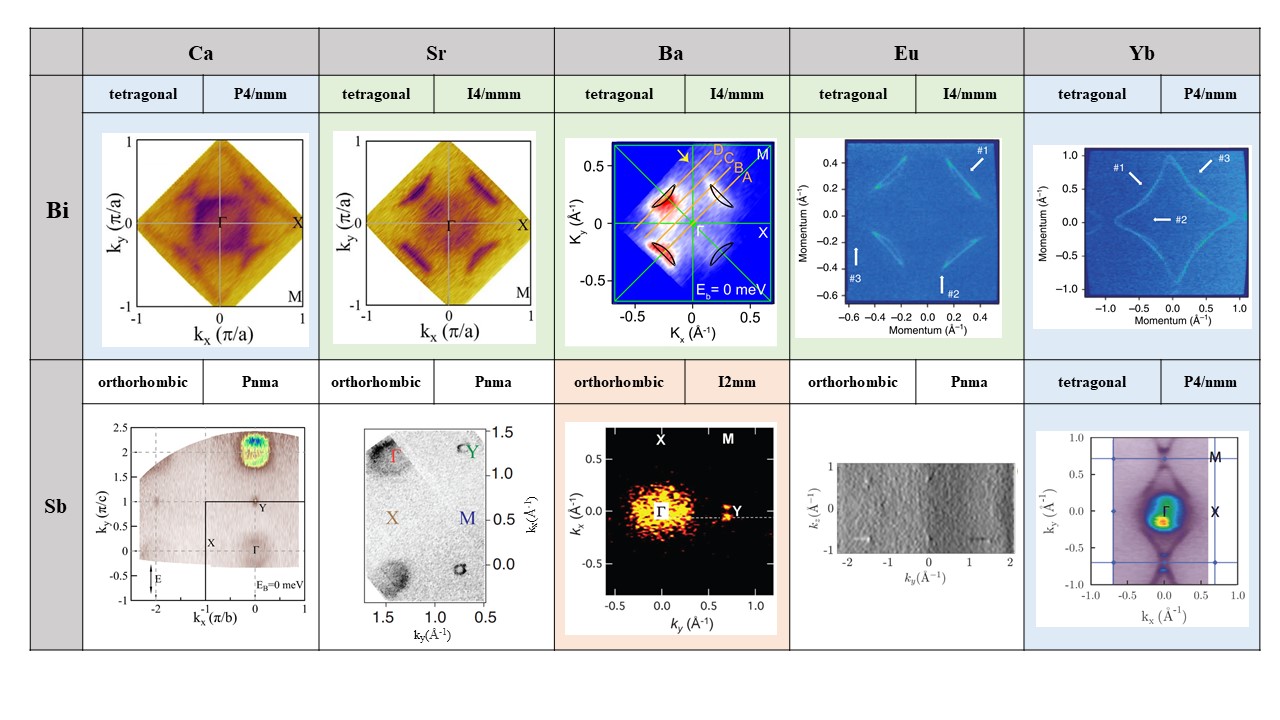}
\end{center}
\begin{center}
\caption{{\bf Summary of crystal structures and electronic structures of AMnPn$_2$ (A=Ca, Sr, Ba, Eu, Yb; Pn=Sb, Bi).} Four kinds of crystal structures are identified with CaMnBi$_2$\cite{YFeng}, YbMnBi$_2$\cite{SBorisenkoYMB} and YbMnSb$_2$\cite{RKealhoferYMS} crystalized in P4/nmm space group; SrMnBi$_2$\cite{JParkSMB,YFeng}, BaMnBi$_2$\cite{HRyuBMB} and EuMnBi$_2$\cite{SBorisenkoYMB} in I4/mmm space group; BaMnSb$_2$\cite{HSakaiBMS,JYLiuBMS} in I2mm space group; CaMnSb$_2$, SrMnSb$_2$\cite{SVRamankuttySMS} and EuMnSb$_2$\cite{JRSohEMS} in Pnma space group. The representative measured Fermi surface mapping for each compound is displayed.
}
\end{center}
\end{figure*}
\end{document}